# Moderate Positive Spin Hall Angle in Uranium

Simranjeet Singh[1], Marta Anguera[1], Enrique del Barco[1*], Ross Springell[2] and Casey W. Miller[3*]

[1]Department of Physics, University of Central Florida, Orlando, Florida USA, 32816

[2]H. H. Will Laboratory, University of Bristol, Bristol BS2 8BS, United Kingdom

[3]School of Chemistry and Materials Science, Rochester Institute of Technology, Rochester, New York, USA, 14623

**Abstract:** We report measurements of spin pumping and the inverse spin Hall effect in $Ni_{80}Fe_{20}$/Uranium bilayers designed to study the efficiency of spin-charge interconversion in a super-heavy element. We employ broad-band ferromagnetic resonance on extended films to inject a spin current from the $Ni_{80}Fe_{20}$ (permalloy) into the uranium layer, which is then converted into an electric field by the inverse spin Hall effect.  Surprisingly, our results suggest a spin mixing conductance of order $2\times10^{19}\,m^{-2}$ and a positive spin Hall angle of 0.004, which are both merely comparable to those of several transition metals. These results thus support the idea that the electronic configuration may be at least as important as the atomic number in governing spin pumping across interfaces and subsequent spin Hall effects.  In fact, given that both the magnitude and the sign are unexpected based on trends in d-electron systems, materials with unfilled f-electron orbitals may hold additional exploration avenues for spin physics.

[*] Authors to whom correspondence should be addressed. *Electronic mails:* delbarco@ucf.edu, cwmsch@rit.edu

In the field of spintronics, one of the major challenges is to create efficient net spin polarization of charge carriers (*i.e.*, spin current) [1]. Out of several available techniques [2-4], dynamical spin pumping [5] is unique because no net charge current is employed, and spin injection can occur over large areas. This has been demonstrated in a wide variety of materials, including transition metals [6-9], organic materials [10] and two-dimensional crystals [11-13], among others. In spin pumping, the precessing magnetization of an externally excited ferromagnet (FM) undergoing ferromagnetic resonance (FMR) is dynamically coupled to the charge carriers in an adjacent non-magnetic system, resulting in a net transfer of spin angular momentum across the ferromagnet/non-magnetic (FM/NM) interface [14]. Experimentally, the pumped spin current can be detected as an enhancement of the Gilbert damping in the ferromagnet due to the loss of angular momentum in every resonant cycle. The injected spin current can be converted into an electrical signal by the inverse spin Hall effect (ISHE) for non-magnetic layers with sufficiently large spin-charge current interconversion [15, 16]. The latter has a figure of merit known as the spin Hall angle, $\theta_H$, defined as the ratio of the inject spin current and resulting charge current. The resultant electromotive force (E$_{ISHE}$) is given by $E_{ISHE} \propto \theta_H J_S \times \sigma$, where the spin current density ($J_s$) is normal to the FM/NM interface, and its polarization ($\sigma$) is dictated by the direction of the ferromagnet's magnetization.

Note that the electrical detection of spin currents via the ISHE depends directly on the magnitude of spin-orbit coupling in the NM layer, as this is an essential interaction to large $\theta_H$. Indeed, barring a few reports [11-13, 17], most spin pumping and ISHE studies concentrate on metals with high spin-orbit coupling. It is generally understood that the spin-orbit coupling, and consequently the spin Hall angle, scales as $Z^4$ (being Z the atomic number) [8, 9, 18]. However, a recent systematic study with transition metals shows that the spin Hall angle can be comparable to that obtained in some heavy metals [9]. This indicates that the electronic structure is quite important in determining spin Hall conductivities and spin Hall angles, leaving the ultimate importance of the atomic number in doubt. Therefore, exploring higher atomic number elements in the periodic table is an important direction for understanding the underlying mechanisms governing the spin Hall effect in non-magnetic materials. As pointed out by Tanaka *et al*. [8, 9, 18], the higher magnetic moment and the smaller band splitting near the Fermi surface of *f*-electron systems when compared to *d*-electron elements, place Lanthanides and Actinides as excellent candidates to display "giant" spin Hall and orbital Hall effects. In this context, this Letter reports the experimental realization of spin pumping and subsequent electrical detection of the generated spin current in the actinide metal Uranium, a super-heavy *f*-electron metal whose atomic number (Z = 92) is larger than any element studied to date in the context of spintronics.

Thin layers of permalloy (Py; $Ni_{80}Fe_{20}$) and bilayers of permalloy/uranium (Py/U) were deposited using a dedicated actinide sputter deposition chamber [19]. DC magnetron sputtering was employed to synthesize samples with thicknesses, $t_{Py}$ and $t_U$ of 12.5 nm and 3 nm, respectively. 3 nm of Nb was used as a capping layer to prevent oxidation of the underlying films. The growth rates were held between 0.05 and 0.1 nm/s in an argon pressure of $7 \times 10^{-3}$ mbar at a temperature of 300 K. The films were deposited onto BK7 glass with rms surface < 0.5 nm. Depleted uranium sputtering targets are commercially available.

Microwave measurements used a broadband on-chip micro-co-planar waveguide (μ-CPW), shielded with a thin insulating polymer layer to prevent interference with the spin pumping process of the sample under study. Extended Py/U thin films were placed upside-down atop the central constricted part of the μ-CPW to achieve maximum signal to nose ratio [20]. The transmission absorption spectra were measured for frequencies up to 18 GHz with the external dc magnetic field applied in-plane. The observed frequency dependence of the FMR linewidth for the Py and Py/U samples is shown in Fig. 1.

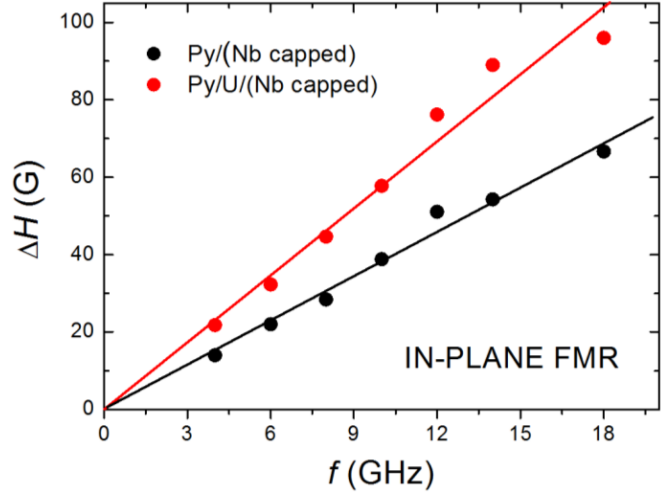

*Figure 1:* *Frequency dependence of the FMR linewidths for Py/Nb and Py/U/Nb samples.*

The presence of Uranium substantially increases the FMR linewidth at all frequencies, while it does not alter the quality of the Py film (as inferred from the linear behavior of the data and the zero-width frequency intercepts). This implies an enhancement of the magnetic damping of Py film due to spin pumping into the U layer. The damping is determined via the slope of the linewidth with frequency, as per the phenomenological Gilbert damping model [21]:

$$\delta H = \delta H_0 + \frac{4\pi\alpha}{\sqrt{3}\,\gamma} f, \qquad (1)$$

where $\gamma = g\,\mu_B/h$ (= 2.88 GHz/kG) is the gyromagnetic ratio and $\alpha$ is the damping parameter, which is related to the Gilbert damping as $G = \alpha\gamma M_s$, with $M_s$ (= 9.5 kG) being the saturation magnetization (the data given in parentheses are obtained from the dependence of the magnetic field position of the FMR peak with frequency, which is not shown here). The first term in Eqn. (1) is the frequency independent inhomogeneous broadening, which is negligible in the samples studied here. The second term denotes the dynamical damping. The extracted damping parameters for the Py/U sample were $\alpha_{Py/U} = 14.3\times10^{-3}$ & $G_{Py/U} = 0.384$ GHz, 150% larger than those of the control sample ($\alpha_{Py} = 9.6\times10^{-3}$ & $G_{Py} = 0.257$ GHz). Note that isolated Py typically has $\alpha_{Py}$ of the order $8\times10^{-3}$ and $G_{Py} = 0.21$ GHz. As we discuss below, the slightly larger values for our Py sample is likely due to the presence of the Nb capping layer.

The observed damping enhancement can be understood as a direct consequence of the spin pumping mechanism which transfers angular momentum from the ferromagnet into the Uranium. The efficiency of spin pumping across the Py/U interface is given by the spin-mixing conductance [21]:

$$g_r = \frac{4\pi M_s d}{\gamma h} (\alpha_{Py/U} - \alpha_{Py}) \qquad (2)$$

where $d$ is the Py thickness. Using the damping parameters obtained from the linewidth-frequency data and the saturation magnetization of the Py, which is obtained from standard FMR measurements, we obtain $g_r = 2.28\times10^{19}$ m$^{-2}$. This value is comparable to those found in Py/heavy-metal interfaces. One needs to keep in mind that Eqn. (2) assumes that no spins flow back into the Py film. This assumption should hold for this thickness of Uranium owing to its large expected spin-orbit coupling and hence small spin diffusion length, which, while unknown at this moment, should be comparable to those found in heavy metals (*e.g.*, 1-4nm in Pt) [22, 23]. This spin mixing conductance value should be taken cautiously since it comes from comparison measurements between Py and Py/U samples, both capped with a 3nm Nb layer. The Nb already contributes to the magnetic damping in the Py/Nb control sample, which enhances the damping slightly from that of isolated Py, as mentioned above. Consequently, using the damping change observed between Py/Nb and Py/U/Nb samples would underestimate the spin-mixing conductance of the Py/U interface. Therefore, $g_r = 2.28\times10^{19}\,\text{m}^{-2}$ should be taken as a lower bound for Py/U. If we use a typical $\alpha_{Py}$ of $8\times10^{-3}$ for isolated permalloy, we estimate $g_r = 3\times10^{19}\,\text{m}^{-2}$.

Next, we turn our attention to the experimental realization of spin-charge conversion in Uranium. For carrying out these measurements two contacts separated a distance of 1 mm were patterned at the surface of the Py/U film using conducting epoxy. The sample was placed on the bottom plate of a high quality factor cylindrical resonant cavity. The electrodes were connected to the external voltmeter by thin copper wires through a small hole in the center of the cavity plate. The resonant cavity is inductively coupled to the external rf source by a small coupling loop performed at the protruding end of a semi-rigid copper coaxial cable attached to the top cavity plate.

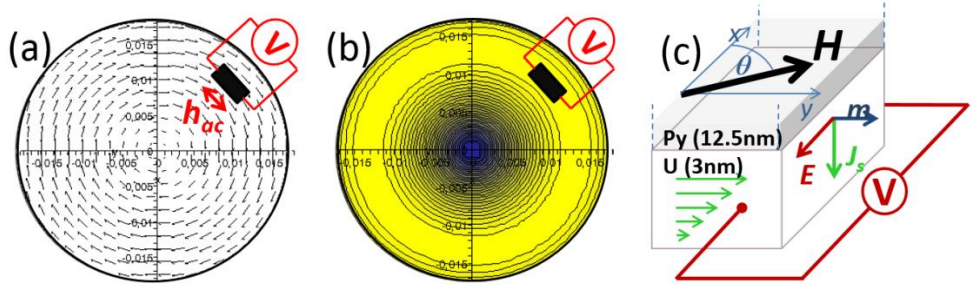

*Figure 2: Schematics of the resonant cavity for microwave excitation and experimental set-up. a) The arrows represent the direction and magnitude (length) of the microwave magnetic field at the bottom plate of the cavity. This configuration corresponds to the TM $_{001}$ mode. b) Color-coded plot of the magnitude of microwave magnetic field at the bottom plate of the cavity. c) The schematic of ISHE experiment. θ is the direction of the externally applied dc magnetic field. The spin currents are injected into uranium with polarization along the y-axis. The spin gradient normal to the interface results in electromotive force along the x-axis.*

The ISHE measurements were carried out by tuning the cavity to its $TM_{011}$ mode (8.54 GHz). The position of the sample within the bottom plate in relation to the direction and magnitude of the rf magnetic field for this mode are illustrated in Figs. 2a and 2b. The sample is placed such that the rf field magnitude is maximum and directed along the direction across which the ISHE electric field will be measured. This configuration helps to minimize the electric field components of the oscillating field along the ISHE field direction. The rf field was pulse-modulated at a few kHz, and the generated ISHE voltage measured by a lock-in amplifier. The schematic of a typical ISHE measurement is depicted in Figure 2c, wherein the precessing magnetization of the Py (about the y-axis) adiabatically pumps a dc component of spin current into the U.

(3)

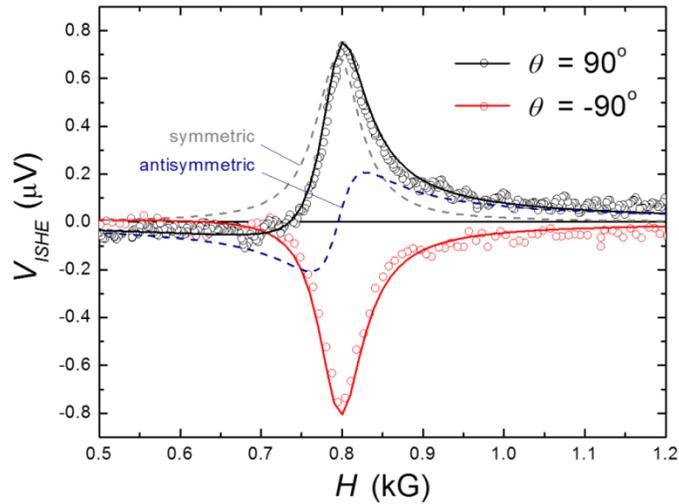

*Figure 3:* *Field dependence of the measured dc voltage in Py/U (circles). The continuous lines are fits to a sum of symmetric and antisymmetric functions (shown for 90 degrees as dashed grey and blue lines, respectively).*

Figure 3 shows the measured dc signal corresponding to the FMR resonance of the sample with the dc magnetic field along opposite directions along the y-axis ($\theta$ = 90 and -90 degrees). Since the spin Hall angle depends on the magnitude of measured ISHE voltage signal, it is necessary to separate the ISHE signal from any other unwanted voltage generation effect. One of the most common alternative voltage contributions can be originated by the anisotropic magnetoresistance (AMR), which is due to a classical induction effect of in-plane oscillating magnetic fields within the film plane [7, 23]. Fortunately, the AMR voltage is antisymmetric around the resonance field [7], contrary to the symmetric nature of the voltage generation expected from the ISHE, as reported in different experimental studies carried out in several FM/NM bilayer systems [7, 16, 17]. In order to separate the ISHE voltage signal arising from the spin pumping in our experiments, the observed signal (black circles in Fig. 3) is fitted (solid black line) using a combination of symmetric, $V_S(H)$, and antisymmetric, $V_{AS}(H)$, functions, as follows:

$$V(H) = V_S(H) + V_{AS}(H) \tag{3a}$$

$$V_S(H) = V_S \frac{\Gamma^2}{(H-H_{res})^2 + \Gamma^2} \;; V_{AS}(H) = V_{AS} \frac{[-2\Gamma(H-H_{res})]}{(H-H_{res})^2 + \Gamma^2} \tag{3b}$$

where $H_{res}$ is the FMR resonance field and $\Gamma$ is the resonance width extracted from the FMR measurements. The symmetric and antisymmetric contributions to the observed data are shown in Figure 3 with gray and blue dashed lines, respectively. The separation between asymmetric and symmetric contributions allows the extraction of the pure ISHE voltage originated from the spin pumping mechanism, which can be extracted from the peak value of the symmetric component at resonance. Figure 3 also shows the switching of the induced voltage upon reversal of the magnetic field polarity, thus confirming the pumping of spins from the permalloy into the uranium film. Perhaps the most conspicuous feature of these results is that the ISHE voltage signal indicates that the spin Hall angle for U is positive.

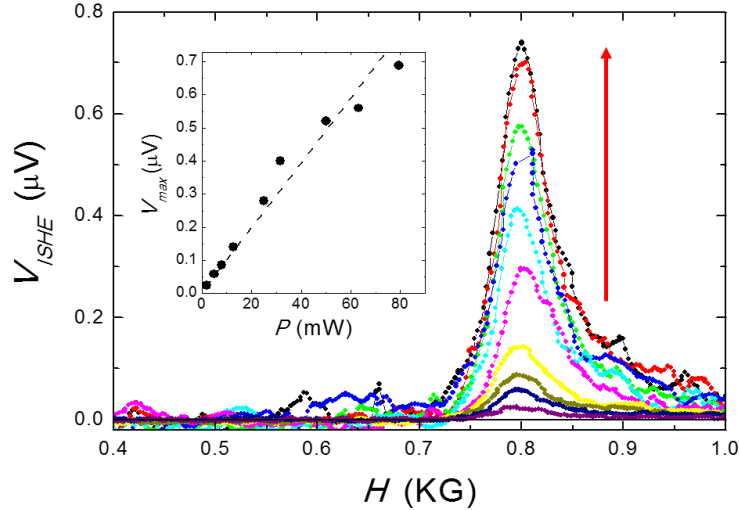

*Figure 4: Microwave power dependence of inverse spin Hall effect voltage signal in Py/U. The red arrow points to the increasing microwave power direction. The inset shows that the amplitude scales linearly with power, confirming the spin pumping origin of the signal.*

According to spin pumping theory, the amplitude of the generated ISHE voltage depends on the microwave power proportionally to the cone-angle of magnetization precession in the ferromagnet. As expected, the measured ISHE signal for increases with microwave power, as shown in Fig. 4. The inset to Fig. 4 shows the power dependence of the amplitude of the ISHE voltage, as extracted from the symmetric component of the signal. The linear behavior of the ISHE voltage with microwave power agrees well with the expectation from spin pumping.

Finally, in order to quantify the spin-charge current interconversion efficiency in uranium we proceed to calculate the spin Hall angle. To write the electromotive force generated in uranium by the ISHE mechanism, we follow the expression given by Ando *et al*. [6]:

$$V_{ISHE} = \frac{w\,\theta_H \lambda_{SD} \text{Tanh}[d_N/2\lambda_{SD}]}{d_N \sigma_N + d_F \sigma_F} \left(\frac{2q}{\hbar}\right) \frac{g_r\, \gamma^2\, h^2 \hbar (M_s\, \gamma + \sqrt{(M_s^2 \gamma^2 + 4\omega^2)})}{8\pi\, \alpha_{Py/U}^2\, (M_s^2 \gamma^2 + 4\omega^2)} \qquad (4)$$

where, $q$ is the electron charge, $\sigma_N = 3.5 \times 10^6 (\Omega m)^{-1}$ is the conductivity of the uranium, $d_N = 3$ nm is the thickness of the uranium film; $\sigma_F = 1.5 \times 10^6 (\Omega m)^{-1}$ is the conductivity of Py, $d_F = 12.5 \times 10^{-9}$ m is the thickness of the Py film, $w = 1 \times 10^{-3}$ m is the separation between contacts, $\omega = (2\pi)f$ is the microwave angular frequency, $M_s = 0.95$ T is the saturation magnetization of Py, g = 2.06 is the Lande factor for Py, $\alpha_{Py/U} = 0.143$ is the damping parameter of Py/U, $g_r = 2.28 \times 10^{19}\, m^{-2}$ is the spin mixing conductance at the Py/U interface, $h = 0.86 \times 10^{-4}\, T$ is the magnitude of rf magnetic field at the sample location, and $V_{ISHE} = 0.76 \times 10^{-6}$ V is the amplitude of the symmetric part of the measured signal at resonance. Since no data are available for the spin diffusion length in uranium, we have used $\lambda_{SD} = 3$ nm (a value within the range of those found in Pt), noting that the obtained result from Eqn. (4) would not change much for $\lambda_{SD} > 3$ nm. In addition, we note that the Nb capping layer may contribute to the ISHE signal if not all spins are diffused within the 3nm-thick U layer, which will ultimately depend on the spin diffusion length of uranium. Due to the negative spin Hall angle for Nb (-0.0087) [16], the small fraction of spin current interconverted to electric current in the Nb film would decrease the signal obtained from U, and lead to an underestimation of the spin Hall angle of the latter, although this effect is expected to be minimal for a 3 nm-thick U film, since most spins would be absorbed before reaching the Nb layer. Using Eqn. (4) with the values given above we find the spin Hall angle in uranium to be $\theta_H = 0.004\,(0.4\%)$. Were we to use the spin mixing conductance estimated relative to isolated Py, $\theta_H$ would be about 30% lower, with all else constant. Surprisingly for such a high atomic number material, this value is lower than that of Pt, the most common system explored for spin pumping/ISHE experiments and for which there is still debate and different values ranging from 0.006 to 0.4 have been reported in the literature [7, 23-25].

According to the rule of thumb that the spin-orbit interaction is proportional to the atomic number, platinum (Z = 78) should present a lower spin-charge interconversion efficiency than uranium. This is contrary to our observations, which points to alternative explanations for the origin of spin-orbit coupling in metallic systems. For example, Du *et al.* have recently reported a comprehensive study of spin pumping at FM/NM interfaces using several transition metals, including Ti (Z = 22), V(Z = 23), Cr(Z = 24), Mn(Z = 25), Ni(Z = 28) and Cu(Z = 29) [9]. They found that the spin Hall angle in some of these elements is comparable to that found in substantially heavier metals (*e.g.*, Pt), and that its value and sign do not follow a proportional law with the atomic weight within the studied series of elements. Indeed, the authors find the spin Hall angle sign to oscillate within the series, being maximum for Cr ($\theta_H = -0.05$) and Ni ($\theta_H = 0.05$), and associate this behavior to a dominant role of the *d*-electron configuration in the spin Hall effect in 3*d* elements. More specifically, these authors find that the effects of atomic number and *d*-orbital filling are additive. For example, for elements with filled electronic orbitals, such as in the Cu, Au, Ag series, the atomic number is the most relevant parameter governing spin-orbit coupling because of their zero orbital moment [8, 9, 18], while for transition metals with partially filled *d*-orbitals, the orbital moment contribution is dominant [8, 9, 18]. It is likely that the same construction applies to Lanthanides (4*f*) and Actinides (5*f* metals), and that the *f*-orbital filling in elements of these series may govern the spin Hall conductivities. Our results show unexpected spin Hall behaviors: aside from the moderate magnitude of the spin-Hall angle in uranium, its positive sign is incommensurate with results of transition metals. Since its electronic configuration ([Rn]$5f^3 6d^1 7s^2$) has less than half filling of both the *d* and *f* orbitals, one would have expected a negative spin Hall angle, as is observed in transition metals. Perhaps, the strong spin-orbit coupling in lanthanides and actinides mixes the spin and orbital Hall effects in the same way that it makes only the total angular momentum, $J = S+L$, to be a good quantum number, and one should start thinking about the total angular momentum Hall effect (JHE) to understand spintronics effects in *f*-electron systems. A comprehensive study involving more elements on these series would be necessary to completely understand the underlying physics behind the spin and orbital Hall effects in heavy metals.

**Acknowledgments**

Work at UCF was supported by NSF-ECCS grant # 1402990. Work at RIT was supported by NSF-ECCS Grant 1515677.